\documentclass[sigconf]{acmart}

\usepackage{booktabs} 
\newcommand{\model}{SUMMON\xspace}

\usepackage[ruled]{algorithm2e} 

\SetAlFnt{\small}
\SetAlCapFnt{\small}
\SetAlCapNameFnt{\small}
\SetAlCapHSkip{0pt}

\acmJournal{TOG}

\definecolor{MyDarkBlue}{rgb}{0,0.08,1}
\definecolor{MyDarkGreen}{rgb}{0.02,0.6,0.02}
\definecolor{MyDarkRed}{rgb}{0.8,0.02,0.02}
\definecolor{MyDarkOrange}{rgb}{0.40,0.2,0.02}
\definecolor{MyPurple}{RGB}{111,0,255}
\definecolor{MyRed}{rgb}{1.0,0.0,0.0}
\definecolor{MyGold}{rgb}{0.75,0.6,0.12}
\definecolor{MyDarkgray}{rgb}{0.66, 0.66, 0.66}
\definecolor{MyWineRed}{rgb}{0.694,0.071, 0.149}
\definecolor{nicegreen}{rgb}{0.1, 0.6, 0.2}

\copyrightyear{2022}
\acmYear{2022}
\setcopyright{rightsretained}
\acmConference[SA '22 Conference Papers]{SIGGRAPH Asia 2022 Conference Papers}{December 6--9, 2022}{Daegu, Republic of Korea}
\acmBooktitle{SIGGRAPH Asia 2022 Conference Papers (SA '22 Conference Papers), December 6--9, 2022, Daegu, Republic of Korea}\acmDOI{10.1145/3550469.3555426} \acmISBN{978-1-4503-9470-3/22/12}

\begin{document}
\title{Scene Synthesis from Human Motion}


\settopmatter{authorsperrow=4}

\author{Sifan Ye}
\affiliation{
    \institution{Stanford University}
    \country{United States of America}
}
\email{sifan.ye@cs.stanford.edu}
\authornote{~and $\dagger$ indicate equal contribution. \url{https://sites.google.com/stanford.edu/summon}}

\author{Yixing Wang}
\affiliation{
    \institution{Stanford University}
    \country{United States of America}
}
\email{yiw998@stanford.edu}
\authornotemark[1]

\author{Jiaman Li}
\affiliation{
    \institution{Stanford University}
    \country{United States of America}
}
\email{jiamanli@stanford.edu}

\author{Dennis Park}
\affiliation{
    \institution{Toyota Research Institute}
    \country{United States of America}
}
\email{dennis.park@tri.global}

\author{C. Karen Liu}
\affiliation{
    \institution{Stanford University}
    \country{United States of America}
}
\email{karenliu@cs.stanford.edu}

\author{Huazhe Xu}
\affiliation{
    \institution{Stanford University}
    \country{United States of America}
}
\email{huazhexu@stanford.edu}
\authornotemark[2]

\author{Jiajun Wu}
\affiliation{
    \institution{Stanford University}
    \country{United States of America}
}
\email{jiajunwu@cs.stanford.edu}
\authornotemark[2]

\begin{abstract}
Large-scale capture of human motion with diverse, complex scenes, while immensely useful, is often considered prohibitively costly. Meanwhile, human motion alone contains rich information about the scene they reside in and interact with. For example, a sitting human suggests the existence of a chair, and their leg position further implies the chair's pose. In this paper, we propose to synthesize diverse, semantically reasonable, and physically plausible scenes based on human motion. Our framework, \textbf{S}cene Synthesis from H\textbf{UM}an \textbf{M}oti\textbf{ON}~(\model), includes two steps.
It first uses ContactFormer, our newly introduced contact predictor, to obtain temporally consistent contact labels from human motion. Based on these predictions, \model then chooses interacting objects and optimizes physical plausibility losses; it further populates the scene with objects that do not interact with humans. Experimental results demonstrate that \model synthesizes feasible, plausible, and diverse scenes and has the potential to generate extensive human-scene interaction data for the community. 
\end{abstract}

%
%


%
%


\begin{teaserfigure}
  \includegraphics[width=\textwidth]{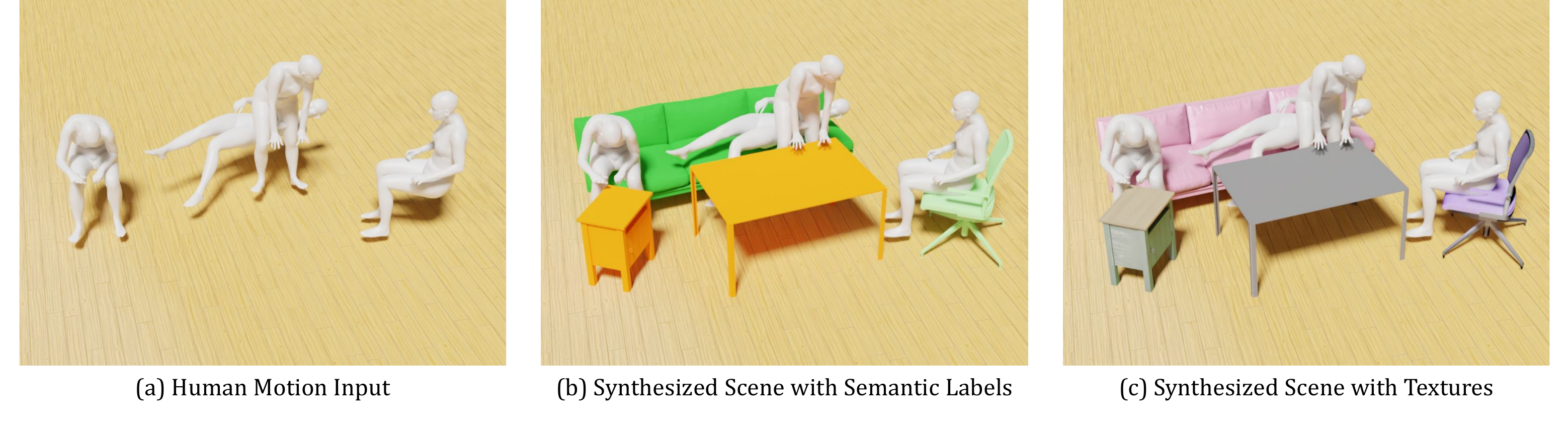}
  \vspace{-8mm}
  \caption{From a human motion sequence, SUMMON synthesizes physically plausible and semantically reasonable objects. }
  \vspace{2mm}
  \label{fig:teaser}
\end{teaserfigure}

\maketitle
\section{Introduction}

Capturing, modeling, and synthesizing realistic human motion in 3D scenes is crucial in a spectrum of applications such as virtual reality, game character animation, and human-robot interaction. To facilitate research in this area, a plethora of datasets~\citep{prox,AMASS,savva2016pigraphs} have been curated to capture human motion. 
For example, \citet{bhatnagar22behave} collected trajectories of humans manipulating objects. The PROX-E dataset~\citep{prox-e} contains human contact with a scene mesh. 
However, building high-quality, large-scale datasets annotated with both diverse human motions and rich 3D scenes remains challenging. This is mainly because current data capture pipelines depend on costly devices, such as MoCap systems, structure cameras, and 3D scanners, and therefore can only be conducted in laboratory settings, which entails limited physical space and scene diversity. Inspired by recent advances in modeling 3D human poses and their contact with environments, we aim to address these challenges by exploring a new possibility: \emph{can we learn to synthesize the scenes only from human motion?} If successful, our system will also have many potential applications beyond data collection, such as providing suggestions during the creation of virtual environments based on artists' motions in VR. 

Recent works have proposed to estimate room layouts based on human trajectories and learned room priors~\citep{nie2021pose2room}. However, only semantics, not affordances, was considered in the reconstructed layouts. \citet{rec-visual} proposed to reconstruct scene objects from visual inputs and then use Human-Scene Interactions (HSIs) to further improve the feasibility. While such a method produces physically plausible reconstructions, it requires additional visual inputs so that the reconstructed scenes are restricted. 

We propose \textbf{S}cene Synthesis from H\textbf{UM}an \textbf{M}oti\textbf{ON} (\model), a method that predicts feasible object placements in a scene based solely on 3D human pose trajectories, as shown in Figure~\ref{fig:teaser}. \model consists of two modules: a human-scene contact prediction module and a scene synthesis module. The human-scene contact prediction module, named ContactFormer, leverages existing HSI data to learn a mapping from human body vertices to the semantic label of the objects that are in contact. ContactFormer advances previous methods~\citep{POSA} by incorporating temporal cues to enhance the consistency in label prediction in time. Given the estimated semantic contact points, the scene synthesis module first searches for objects that fit the contact points in terms of semantics and physical affordances to the agent; it then populates the scene with other objects that have no contact with humans, based on human motion and objects inferred from previous steps.

We conduct our experiments using the PROXD~\citep{prox} and the GIMO~\citep{zheng2022gimo} datasets. In terms of contact estimation, ContactFormer outperforms previous single-frame contact prediction methods~\citep{POSA}. In terms of scene synthesis, our proposed system shows more realistic, physically plausible, and diverse scenes than baselines, using various metrics and human evaluation. 

Our contributions are threefold. First, we propose \model, a system that synthesizes semantically reasonable, physically plausible, and diverse scenes based only on human motion trajectories. 
Second, as a part of \model, we propose a contact prediction module ContactFormer that outperforms existing methods by modeling the temporal consistency in semantic labels. 
Third, we demonstrate that the scenes synthesized by \model consistently outperform existing methods both qualitatively and quantitatively. 

\section{Related Works}

\paragraph{Scene affordance learning.}
Learning affordance from human-scene interaction has caught much attention recently~\cite{fouhey2012people, wang2019geometric, chuang2018learning, delaitre2012scene, gupta20113d, chen2019holistic++,zhu2014reasoning}. In the literature, researchers study how to put human skeletons in a scene. For example, \citet{bingewatching} proposed to learn the affordance from sitcom videos for positioning skeletons in a static image. \citet{gen3dpose} introduced a generative model of 3D poses to predict plausible human poses in a scene. 
Along with developing better human body representations, there have been methods that try to put a 3D human body into the scene~\cite{prox-e}. More recently, POSA~\cite{POSA} learns a model that augments a SMPL-X human body model vertices with contact probability and semantic labels to place human poses in a 3D scene mesh. \citet{body-aware-gen} proposed a fitting and comfort-based loss to train an affordance-aware model to generate chairs that fit a human body pose. 
Several works also try to collect or generate data that involve human-scene interactions. For example, VirtualHome~\cite{puig2018virtualhome} provides a simulated 3D environment where humanoid agents can interact with 3D objects. BEHAVE~\cite{bhatnagar22behave} provides a dataset of real full-body human parameterized using SMPL interacting and manipulating objects in 3D with contact points. Our work takes an additional step from the affordance learning works: we first learn to understand the affordance, then leverage them to synthesize scenes that can be used for other related tasks.

\paragraph{Human motion synthesis.}
Motion synthesis is a long-standing problem in computer graphics and vision~\citep{brand2000style, kovar2003flexible, park2002line, spallone2015digital, holden2016deep}.  \citet{xu2020hierarchical} proposed a hierarchical way to generate long-horizon motion by using a memory bank to retrieve short-horizon reference clips. \citet{harvey2020robust} proposed to predict motion robustly with additional embeddings. Recently, many works also take the environment into consideration~\citep{hassan_samp_2021,wang2021synthesizing,rempe2021humor}.
For example, \citet{wang2021synthesizing} combined long-term human motion synthesis conditioned on a scene mesh with affordance optimization to generate realistic human trajectories. SAMP~\cite{hassan_samp_2021} learns generalized interaction for object classes across different instances of that class. Our work is trying to solve the inverse problem that generates plausible scenes given human motion trajectories.

\paragraph{Scene synthesis.}
Our work is also closely related to synthesizing plausible 3D scenes and room layout~\citep{zhou2019scenegraphnet,li2019grains,purkait2020sg,luo2020end,wang2019planit,zhang2020fast,ritchie2019fast,wang2021sceneformer}. For example, ATISS~\citep{ATISS} learns an autoregressive generative model for furniture placement. It can be used for generating plausible novel room layouts, completing a scene given existing objects, and suggesting possible placements given spatial constraints. Another work, Pose2Room~\citep{nie2021pose2room}, predicts bounding boxes of objects conditioned on 3D human pose trajectory. MOVER~\citep{rec-visual} reconstructs 3D objects constrained by 3D human body predictions from monocular RGB videos. Unlike these prior methods, our model generates not only layouts but also affordable objects with only human trajectories.

\section{Method}

\begin{figure*}[ht!]
    \includegraphics[width=7.1in]{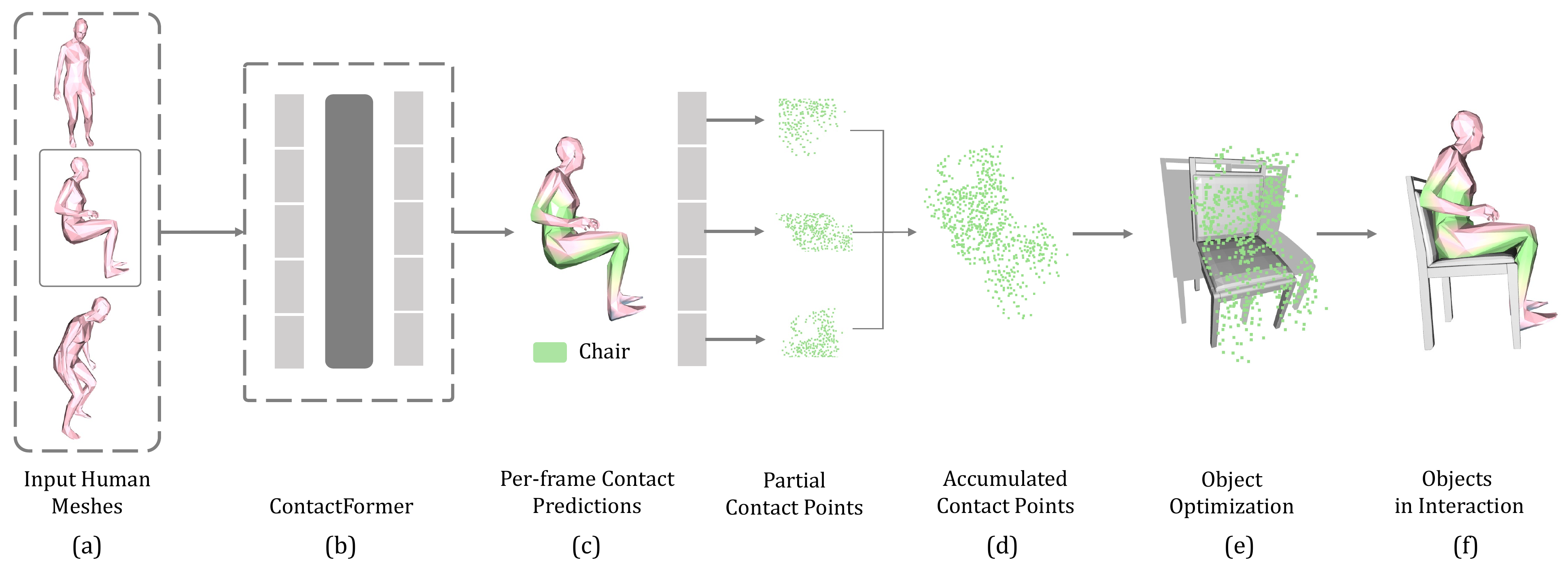}
    \vspace{-9mm}
    \caption{\textbf{The overview of \model}: (a) an input sequence of human body meshes interacting with a scene, (b) the ContactFormer that predicts per-frame contact labels, (c) per-frame contact predictions, (d) estimated contact points, (e) synthesized objects, and (f) objects in interaction. }
    \vspace{-3mm}
    \label{fig:method}
\end{figure*}

We aim to predict a set of furniture objects and a physically plausible 3D configuration of them only from human motion sequences. We first introduce the human body and contact representation in Sec.~\ref{sec:representation}. \model generates a temporally consistent contact semantic estimation for each vertex of the human body to retrieve suitable objects~(Sec.~\ref{sec:ContactFormer}). Then we optimize object placement based on the contact locations and physical plausibility~(Sec.~\ref{sec:obj_rec}). An illustration of our method is shown in Figure~\ref{fig:method}.

\subsection{Human Body and Contact Representation}
\label{sec:representation}
We use a modified version of SMPL-X~\cite{smplx} as the representation of human body poses. Specifically, we parameterize the human body with $M(\theta, \beta): \mathbb{R}^{|\theta| \times |\beta|} \rightarrow \mathbb{R}^{3N}$, where $\theta$ denotes pose parameters, $\beta$ denotes coefficients in a learned shape space, and $N$ is the number of vertices in a SMPL-X body mesh. For computation efficiency, we downsample the vertices from $10{,}475$ to $655$ points, following the prior work by~\citet{POSA}.

We represent contact information by per-vertex features. For each vertex $v_b \in V_b$, where $V_b$ is all vertices of a human body, we use a one-hot vector $f$ to represent the contact semantic label for that vertex. Each vector $f$ has a length of $|f| = C + 1$, where $C$ is the number of object classes. We introduce an extra ``void'' class to represent vertices without contact. We use $F$ to denote the contact semantic labels for all vertices in a body pose.

\subsection{Human-Scene Contact Prediction}
\label{sec:ContactFormer}

\begin{figure}[t]
\begin{center}
    \includegraphics[width=1.0\columnwidth]{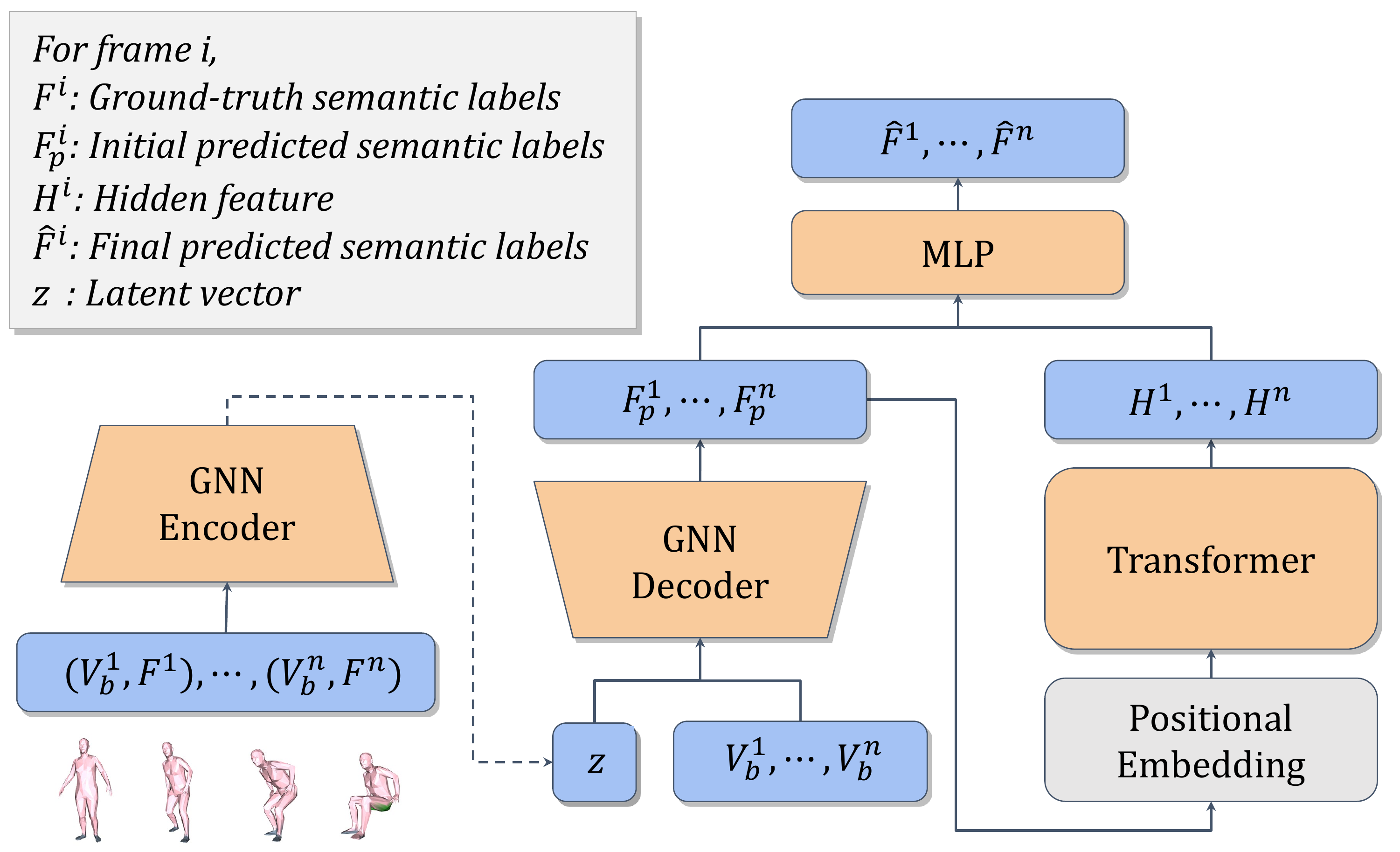}
    \vspace{-5mm}
    \caption{\textbf{The architecture of ContactFormer.} We first use a GNN-based variational autoencoder to encode the contact points. Then a transformer is applied to improve the temporal information fusion. We also add a sinusoidal positional encoding to the output of the GNN decoder.}
    \vspace{-3.5mm}
    \label{fig:network}
\end{center}
\end{figure}

Our dataset consists of a sequence of paired vertices and contact semantic labels $\{ (V^1_b, F^1), (V^2_b, F^2), ...,  (V^n_b, F^n) \}$, where $V^i_b$ represents the human body vertices~(Figure~\ref{fig:method}(a)), $F^i$ represents the contact semantic labels for frame $i$, and $n$ is the varied sequence length. We first train a conditional Variational Autoencoder~(cVAE) to learn a probabilistic model of contact semantic labels conditioned on vertex positions. Then we deploy transformer layers on top of the cVAE to improve temporal consistency. We refer to this framework as ContactFormer. An illustration of the overall network architecture is shown in Figure~\ref{fig:network}.

\paragraph{Contact semantics prediction.} 
We first train a model to predict contact semantic labels for each individual pose. Given a pair of body vertices and contact semantic labels $(V_b, F)$, we first fuse these two components: $I_e = \text{Concat}(V_b, F)$. We feed $I_e$ into a graph neural network~(GNN) encoder $G_{Enc}$ to get a latent Gaussian space with the mean $H_{\mu}$ and the standard deviation $H_{\sigma}$. Then we sample a latent vector $z$ from the latent Gaussian space and concatenate it with each vertex position: $I_d = \text{Concat}(V_b, z)$. We feed $I_d$ into a GNN decoder $G_{Dec}$ to predict the reconstructed contact semantic labels $F_p$. Note that both GNNs in the encoder and the decoder share the same structure as in~\citet{POSA}. Each vertex feature $h^k_x$ for vertex $x$ at layer $k$ is updated by
\begin{equation}
    h^k_x = \text{Linear}(\text{Concat}(\{ h^{k-1}_{x'}: x' \in N(x) \})),
\end{equation}
where $N(x)$ is defined as the $m$-nearest neighbor vertices of $x$ in a spiral-ordered sequence, as proposed by~\citet{spiralnet}.

\paragraph{ContactFormer:} We train a transformer to extract temporal information from a pose sequence to enhance prediction consistency, as shown in Figure~\ref{fig:network}. Specifically, given a sequence of pose and contact semantic labels $\{ (V^1_b, F^1), ..., (V^n_b, F^n) \}$ from frame $1$ to $n$, we first use the previous model to reconstruct contact semantic labels $F^i_p$ independently for each frame $i$. We then embed each $F^i_p$ into a hidden feature space, augmenting it with a sinusoidal positional embedding before feeding it to the transformer module. The output of the transformer module is a sequence of n vectors $\{ H_1, ..., H_n \}$. For each frame $i$, we concatenate $H_i$ with the initial prediction $F^i_p$ and use a multi-layer perceptron (MLP) to get a final prediction $\hat{F}^i$. The final prediction is shown in Figure~\ref{fig:method}(c).

\paragraph{Training:} We optimize the model's parameters by the following loss function:
\begin{equation}
    \mathcal{L} = \mathcal{L}_{rec} + \alpha \cdot \mathcal{L}_{KL},
\end{equation}
where $\mathcal{L}_{rec}$ is the sum of the categorical cross entropy~(CCE) loss between the ground truth semantic label $F^i$ and the model prediction $\hat{F}^i$ for any frame $i$:
\begin{equation}
    \mathcal{L}_{rec} = \sum_i \text{CCE} (F^i, \hat{F}^i),
\end{equation}
and $\mathcal{L}_{KL}$ is the Kullback-Leibler divergence loss between the latent Gaussian space and the normal distribution $\mathcal{N}$:
\begin{equation}
   \mathcal{L}_{KL} = KL(Q(z|F, V_b)||\mathcal{N}). 
\end{equation}
Here we use $Q$ to represent the encoder network in our cVAE combined with the sampling process with the reparameterization trick. Inspired by \citet{beta-vae}, we also multiply $\mathcal{L}_{KL}$ with a weight $\alpha$ to control the balance between the reconstruction accuracy and diversity.

\subsection{Scene Synthesis} \label{sec:obj_rec}
\begin{figure}[t]
\begin{center}
    \includegraphics[width=1.0\columnwidth]{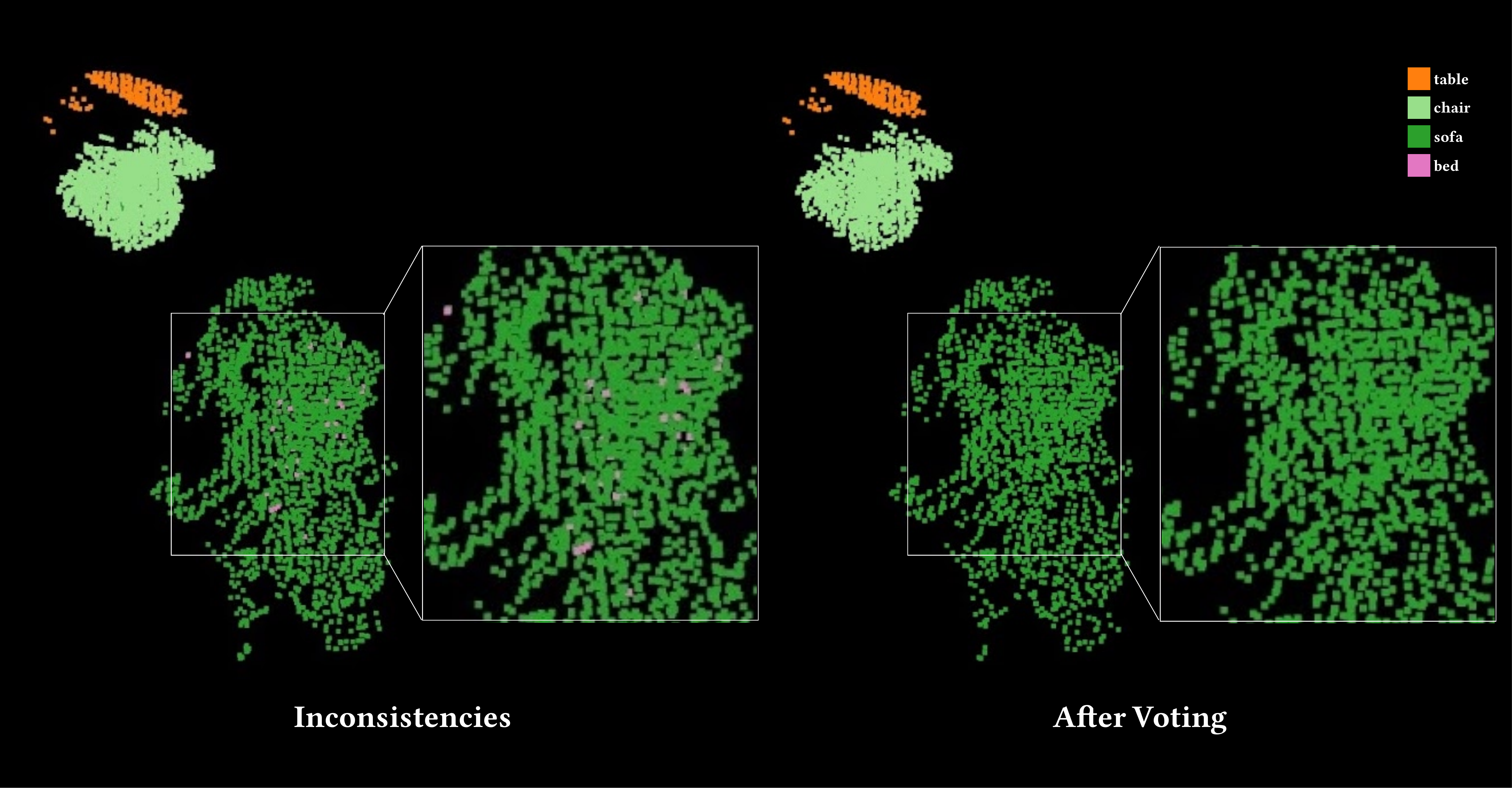}
    \vspace{-5mm}    \caption{\textbf{Illustration of the local majority voting}. From the zoomed-in box, there are multiple inconsistent points in the original contact points. The pink points represent the semantic label bed, and the green points represent the label sofa. We alleviate this issue by adding majority voting.}
    \vspace{-5mm}
    \label{fig:vote_local}
\end{center}
\end{figure}

\paragraph{Contact Object Recovery} Given the accumulated contact points from each frame predicted by ContactFormer~(Figure~\ref{fig:method}(d)), we further reduce spatial prediction noise by performing a local object class majority voting as shown in Figure~\ref{fig:vote_local}. Then, the vertices of each predicted object class are clustered into possible contact instances $V_c$, using the shortest length of all object edges in that class as $\epsilon$ for clustering. In practice, we downsample the contact vertices to keep later computations tractable. 

We then optimize the poses of the object point cloud $V_o$ by minimizing the following losses:
\begin{equation}
    \mathcal{L}(V_c, V_o) = \mathcal{L}_{contact} + \mathcal{L}_{pen}.
    \label{eq:placement_objective_func}
\end{equation}
The contact loss $\mathcal{L}_{contact}$ is defined as
\begin{equation}
    \mathcal{L}_{contact} = \lambda_{contact} \frac{1}{|V_c|} \sum_{v_c \in V_c} \min_{v_o \in V_o} || v_c - v_o||^2_2,
    \label{eq:contact_loss}
\end{equation}
where $\lambda_{contact}$ is a tunable hyperparameter. This loss encourages the object to be in contact with the predicted human vertices.
The penetration loss $\mathcal{L}_{pen}$ is defined as:
\begin{equation}
    \mathcal{L}_{pen} = \lambda_{pen} \sum_{d_c^i < t} {d_c^i}^2,
    \label{eq:pen_loss}
\end{equation}
where $d_c^i$ are signed distances between the object and the human body sequence, $t$ is the penetration distance threshold. This loss prevents the object from penetrating the human body sequence.

Intuitively, these losses encourage objects to be in contact with human meshes, but not penetrate them. An illustration of the optimized object placement is shown in Figure~\ref{fig:method}(e). To improve computation efficiency, we choose to compute human SDF from merged human meshes of the motion sequence. To have a consistent scale of loss across different objects, we choose the number of sampled points according to the size of the object.

\paragraph{Constrained Scene Completion}
\label{sec:scene_completion}
To obtain a complete scene, we also predict non-contact objects as a scene completion task constrained by 3D human trajectories and existing in-contact objects. The floor is divided into a grid, and each cell is labeled as occupied if feet vertices or object vertices are in close proximity.
Considering the furniture categories in a room as a sequence, we train an autoregressive transformer model on the 3D-FRONT dataset~\cite{3dfront}. The model takes as input the categories of existing objects and returns a probability distribution of the next potential object category. We sample a category from the distribution and randomly place an object of that category onto an unoccupied floor grid. To prevent the sampled object from penetrating the human body sequence, we further optimize the object's translation and rotation using our $\mathcal{L}_{pen}$ (see Equation~\ref{eq:pen_loss}).

\section{Experiment Setup}
In this section, we introduce the datasets and implementation details for the whole \model framework.

\subsection{Datasets} \label{dataset}
We use the PROXD~\cite{prox} dataset for training our ContactFormer. PROXD uses RGB-D cameras to capture 20 human subjects interacting with 12 scenes. We represent human poses using the SMPL-X format to reconstruct human body meshes.
The pose sequences in PROXD are estimated using SMPLify~\cite{smplify} and contain many jitters. We apply LEMO~\cite{lemo}, a learned temporal motion smoothness prior, to produce smooth human motion as training data.
Our ground truth per vertex contact semantic labels are generated using scene SDF with contact semantic labels from PROX-E~\cite{prox-e}, which extends PROXD by manually annotating the scene meshes with predefined object categories. We define human-scene contact as the signed distance between a human vertex and the scene to be less than $0.05$.

We select objects from 3D-FUTURE~\cite{3dfuture} to be placed into the scenes. 3D-FUTURE is a dataset of categorized 3D models of furniture with their original sizes. We use a selected subset of 3D-FUTURE to reduce candidate search time. To simplify contact estimation and limit predicted object classes to the available ones in our object dataset, we reduce the contact object categories in the PROX-E dataset from 42 to 8. 

We use the GIMO dataset~\cite{zheng2022gimo} as another test dataset for evaluating the generalization ability of the proposed method on out-of-distribution data.

\subsection{Implementation}

\paragraph{ContactFormer.} \label{sec:ContactFormer_imp}
For the encoder and decoder GNNs, we choose the number of hidden layers to be $3$. The dimension for each hidden vertex feature in the GNNs is 64. In the GNN encoder, we downsample the body vertices after each hidden layer by a factor of $4$. We deploy a similar architecture for the transformer layers as used in the previous work~\cite{transformer}. We provide training details and hyperparameter choices in the supplementary materials.

To compare different architectures' capacities for extracting temporal information, we also implement models that use MLP and LSTM~\citep{lstm} modules as the final block on top of the GNN decoder. For the model that uses the MLP module, we deploy a max pooling layer to the output of the GNN decoder along the dimension of vertices. Then we feed it to an MLP block to get the embedding for the whole sequence. The sequence embedding is then fused with the output of the GNN decoder to get the final prediction via a linear projection. For the model that uses the LSTM module, we linearly project the outputs from the GNN decoder into a higher dimensional embedding space and feed them to a bidirectional LSTM layer to extract features for each frame. Frame features are then concatenated with the output from the GNN decoder to obtain final semantic labels.

\paragraph{Contact Object Recovery.}
To reduce noise in contact semantic estimation, we use majority semantic voting in point cloud clusters with $\epsilon=0.1$ and $minPts = 10$. In point cloud clustering for object instance fitting, we used different values for $\epsilon$ for different classes due to their different sizes.  

To place objects into the scene at an appropriate height, we first cluster all the human body vertices that are in contact with the floor. We then take the minimum medians of all clusters as the estimated floor height. Next, we translate the object to place its lowest vertex on the floor. 

To avoid local minima, we perform a grid search for translation along the floor plane and rotation around the up axis to warm-start the initial transformation. We then optimize for the same transformation parameters on top of the results from the grid search. In both cases, We use different $\lambda_{contact}$, $\lambda_{pen}$ and $t$ to accommodate for different properties of object classes. We keep the transformation that achieves the lowest loss as the optimization result.

To achieve scene diversity, we consider inter-class and intra-class diversity. Inter-class diversity is when a human motion is likely to interact with different classes of objects. For example, sitting down can be performed on a chair, a bed, or a sofa. To achieve this, we first sample per-vertex contact semantics based on the contact probability distribution predicted by ContactFormer. During local clustering of contact object recovery, we consider class labels in local clusters as a probability distribution and sample the cluster contact class. Intra-class diversity is when a human motion is likely to interact with different instances of the same object class. To achieve this, we perform grid search and optimization on all the instances from the object class.

\section{Evaluations}

In this section, we introduce evaluation metrics, baselines, and results on contact prediction and scene synthesis. We encourage the readers to watch the video in the supplementary materials.
\subsection{Contact Semantic Prediction}
\label{sec:csp}
\paragraph{Baselines.} We compare with three baselines, including POSA~\citep{POSA}, an architectural variant that uses a multi-layer perceptron~(MLP) based predictor, and a temporal information fusion variant that uses a bidirectional LSTM~\citep{lstm}.

\paragraph{Metrics.} We use two metrics for evaluating the contact semantic prediction: reconstruction accuracy and consistency score. The reconstruction accuracy is computed as the average correctness of the predicted label compared with the ground-truth label for each vertex. The consistency score is designed following this intuition: if we accumulate predicted contact points from each frame, close contact points should have consistent contact semantic labels. Hence, this loss is computed as follows: Given a pose sequence and the accumulated contact points, for each point, we compare its predicted contact label with the contact labels of its neighboring points to see if the prediction agrees with the majority of the neighboring contact labels (i.e., a high consistency score).

\begin{table}[t]
      \centering
    \caption{\textbf{Results of contact prediction}. We use the reconstruction accuracy and the consistency score as metrics. Our ContactFormer clearly outperforms the baselines.}
    \vspace{-3mm}
    \label{table:contact_comparison}
    \small{
    \begin{tabular}{lccc}
        \toprule
        Models & Reconstruction Acc. $\uparrow$ & Consistency Score $\uparrow$ \\
        \midrule
        MLP Predictor & 0.9082 & 0.8922 \\
        LSTM Predictor & 0.9087 &  0.9209 \\
        POSA    & 0.9106 & 0.8816 \\
        ContactFormer (ours) & \textbf{0.9120} & \textbf{0.9518}\\
     \bottomrule
    \end{tabular}}
    \vspace{-3mm}
\end{table}

\begin{figure}[t]
\begin{center}
    \includegraphics[width=1.0\columnwidth]{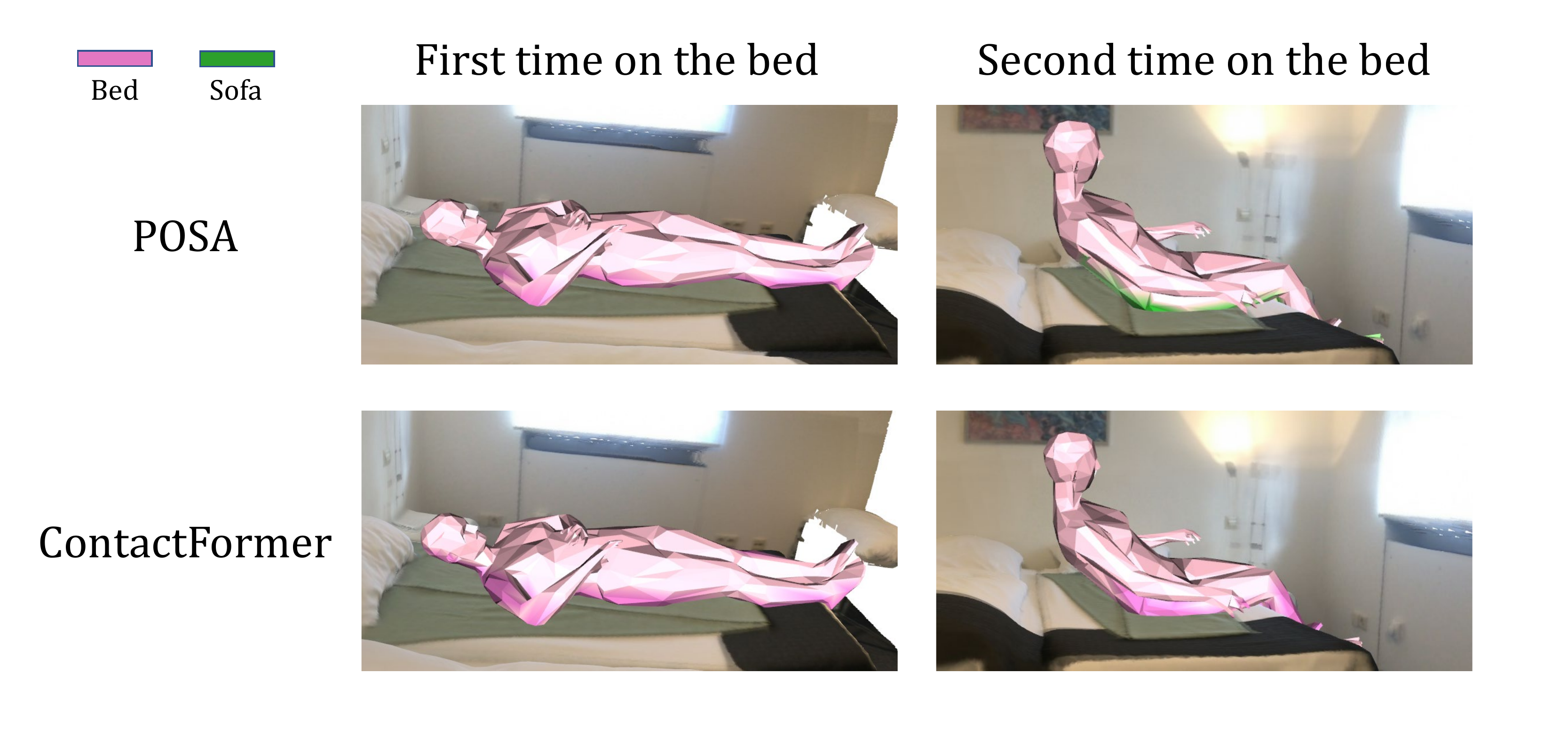}
    \vspace{-8mm}
    \caption{\textbf{Visualizations of the contact prediction results of POSA and our ContactFormer.} Left: Contact predictions from POSA and ContactFormer when the person lies on the bed. Right: Contact predictions from POSA and ContactFormer when the person lies on the bed again after walking around. ContactFormer has better consistency when the person lies in bed for the second time.}
    \vspace{-2mm}
    \label{fig:twobed}
\end{center}
\end{figure}

\begin{figure*}[t]
\begin{center}
    \includegraphics[width=1.0\textwidth]{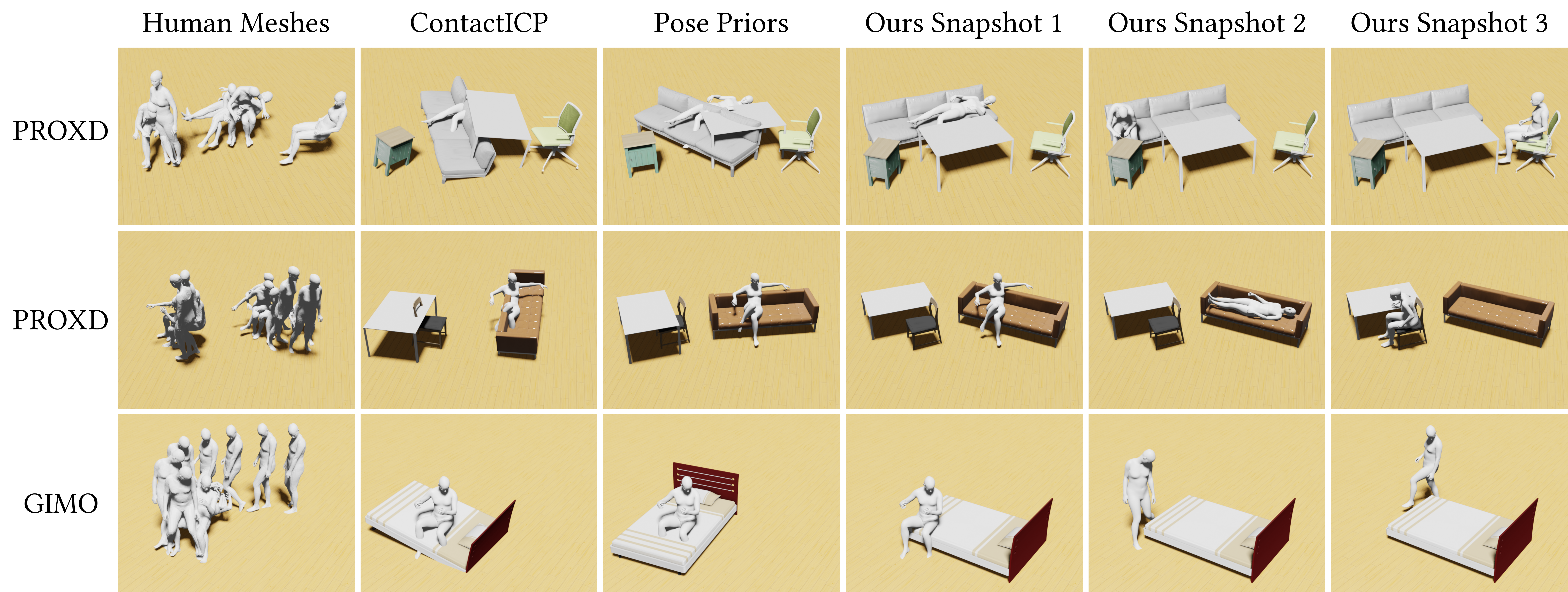}
    \vspace{-5mm}
    \caption{\textbf{Visualizations of the generated objects based on human trajectories}. The human trajectories are from the PROXD dataset and the unseen GIMO dataset. The first column shows the trajectory. The second column shows the results from the ContactICP baseline. The third column shows the results from the Pose Prior baseline. The fourth to sixth columns are snapshots of results generated by \model.}
    \label{fig:baselines}
    \vspace{-5mm}
\end{center}
\end{figure*}

\paragraph{Results.}  Table~\ref{table:contact_comparison} shows the reconstruction accuracy and the consistency score of all methods on the validation set of PROXD. We find that ContactFormer achieves competitive performance in terms of reconstruction accuracies and significantly outperforms all the baselines in consistency scores. This demonstrates the superiority of the transformer-based architecture in predicting temporally consistent yet accurate contact labels.

Figure~\ref{fig:twobed} visualizes the output contact labels from ContactFormer and POSA. We notice that ContactFormer predicts consistent labels the second time the human tries to lie on the bed, while POSA, due to its lack of temporal information, predicts a different label.

\paragraph{User study.}
We conduct a user study to evaluate the quality of the contact semantic label predictions, where we compare ContactFormer with POSA. For each pose sequence in the validation dataset, we render a video showing the human motion, predicted contact semantic labels, and the ground truth scene. We show the predicted contact semantic labels by rendering small areas around body vertices in different colors depending on their labels. Each video is rendered from a camera angle that can clearly capture human motion and semantic labels. For each pose sequence, we ask the human subjects the following question: "Which video seems to have a more reasonable contact label prediction?". Among 22 users, $78.12\%$ of the users choose ContactFormer over POSA, believing ContactFormer provides more reasonable and convincing results. This result echoes the quantitative results in Table~\ref{table:contact_comparison}.


\subsection{Contact Object Recovery}

\paragraph{Baselines.} Since our problem is novel and there are no baselines, we devise two reasonable baselines ourselves: contact-informed point cloud registration~(ContactICP) based on point-to-point ICP~\citep{ICP} and object alignment with pose priors~(Pose Priors) based on the orientation of the hip. We provide the details of those methods in the supplementary materials.

\paragraph{Metrics} 
We use the \textit{non-collision score} proposed by~\citet{prox-e}. This score estimates the collision ratio between human body mesh and scene objects. Since all the methods, including \model, first align the object to the centroid of contact points, contact constraints are naturally satisfied.

\begin{figure*}[t]
\begin{center}
    \includegraphics[width=1.0\textwidth]{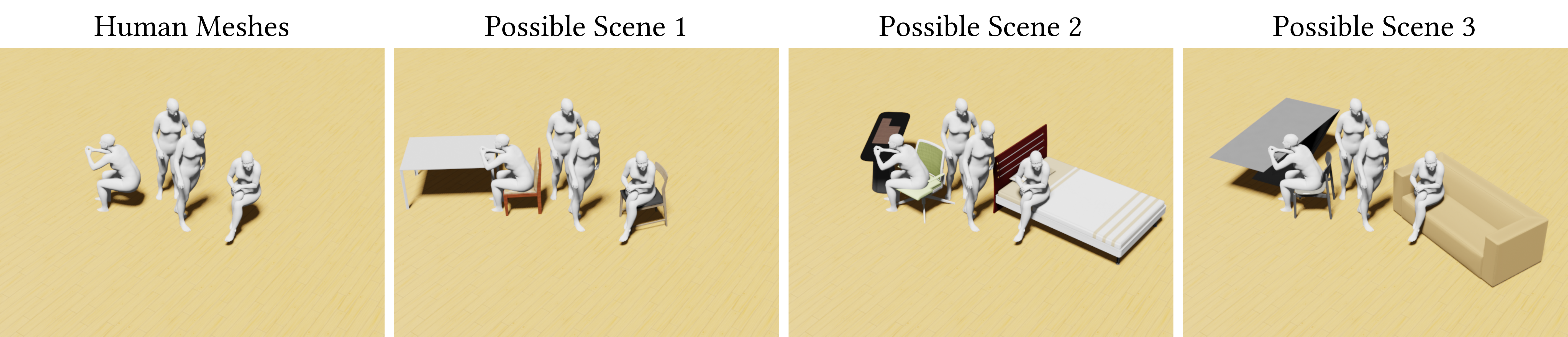}
    \vspace{-7mm}
    \caption{\textbf{Visualizations of possible alternative object placements generated by \model based on the same human trajectories.} In this example, an in-contact object can be a chair, a sofa, or a bed, as long as it does not violate physical constraints. \model can also generate different instances~(e.g., chairs) within the same category.}
    \vspace{-2mm}
    \label{fig:diversity}
\end{center}
\end{figure*}

\begin{figure*}[t]
\begin{center}
    \includegraphics[width=1\textwidth]{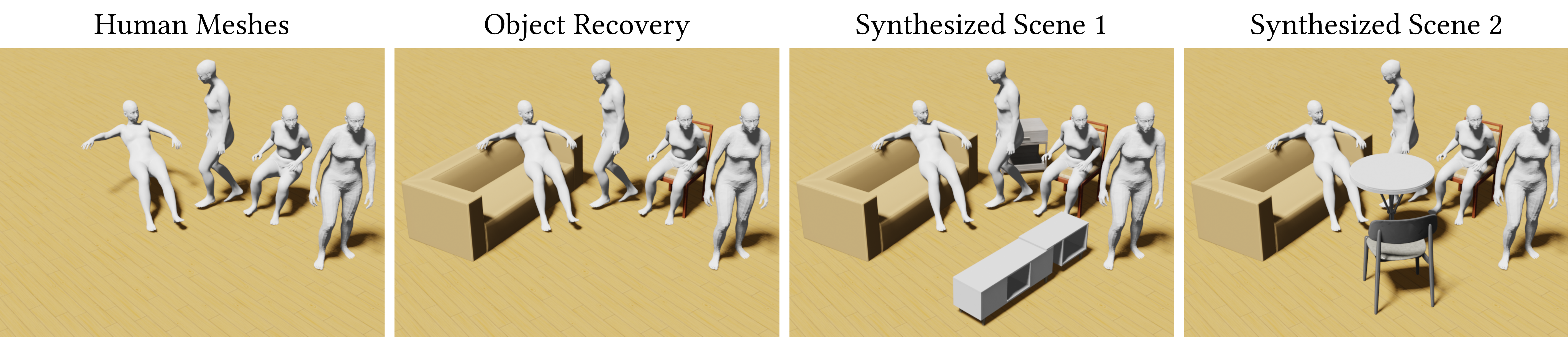}
    \vspace{-5mm}
    \caption{\textbf{Visualizations of scene completion.} Based on all the in-contact objects and human motion trajectories, \model now generates the objects that are not in contact with human meshes. While there is no contact, it makes the scene more complete and introduces the potential for future synthesized human motion sequences to interact with additional objects.}
    \label{fig:completion}
    \vspace{-2mm}
\end{center}
\end{figure*}

\begin{table}[t]
    \centering
    \caption{Non-collision scores for contact object recovery on the smoothed PROXD and the unseen GIMO dataset. For each sequence, the score is computed to be the mean of all possible generated scenes. Higher scores are better.}
    \vspace{-3mm}
    \begin{tabular}{lcc}
        \toprule[0.5mm]
        Method & PROXD & GIMO \\
        \midrule
        ContactICP                         & $0.654$          & $0.820$ \\
        Pose Priors                        & $0.703$          & $0.798$ \\
        \model w/o optimization                 & $0.815$          & $0.937$ \\
        \model (ours) & $\mathbf{0.851}$ & $\mathbf{0.951}$  \\
        \bottomrule
    \end{tabular}
    \vspace{-2mm}
    \label{table:placement_eval}
\end{table}

\paragraph{Results.} 
For each sequence, we compute the mean of the \textit{non-collision scores} for all the objects in the scene. In Table~\ref{table:placement_eval}, we compare the mean non-collision scores on the smoothed PROXD dataset~\cite{lemo}, which was used during training, and the unseen GIMO dataset~\cite{zheng2022gimo}, which also provides SMPL-X parameters for humans interacting with scenes.
 
We visualize comparisons between our method and the baselines in Figure~\ref{fig:baselines}. We find that \model can synthesize objects that are physically plausible and semantically reasonable. ContactICP usually suffers from large penetrations because the contact points might be sparse for registration. While Pose Priors can have seemingly correct object locations and orientations, it often fails to consider physical constraints.

Figure~\ref{fig:diversity} demonstrates various possible scenes generated from the same human motion trajectory by \model. We find that \model can generalize intra-class~(e.g., chairs with different appearances) and inter-class~(e.g., sofa to a bed). We provide additional examples in the supplementary materials.

\paragraph{Human user study.} We follow the same procedure as in Section~\ref{sec:csp}. Instead of contact prediction, we present the users with the animated human motion sequences and the predicted objects in the scene, and ask them to choose the most plausible and realistic placement. 
 From the statistics, we find that $74.5\%$ of the users select \model over ContactICP and Pose Priors.  We find that Pose Priors has a $23.5\%$ user selection rate, showing that it can produce reasonable results in some cases.

    

\paragraph{Ablations.} We also perform ablation on the optimization objectives. Table~\ref{tab:ablation} shows that both the penetration loss and the contact loss are important for \model. Intuitively, the penetration loss helps the object to avoid a collision, while the contact loss helps to keep the object close to humans. We use both the \textit{non-collision score} and the \textit{contact score}. The \textit{contact score} is computed as the fraction of objects in the scenes that are in contact with the human trajectory~\citet{prox-e}.

\subsection{Scene completion}
To generate a full-fledged scene, we train another object generation model following~\citet{ATISS} as in Section~\ref{sec:scene_completion}. The model outputs a family of possible objects that does not contact or penetrate human meshes. Using this model, we generate a fuller scene with both in-contact and no-contact objects. Visualized results are in Figure~\ref{fig:completion}. The completed scenes have additional objects, such as a TV stand or a coffee table. While there is no contact between these objects and the human meshes, they make the scene semantically more realistic. 

\begin{table}[t]
    \centering
    \caption{\textbf{Ablation study on the losses}. The penetration loss and the contact loss are ablated. We use the non-collision score and the contact score as metrics.}
    \vspace{-3mm}
    \begin{tabular}{lcc}
        \toprule[0.5mm]
        Method & non-collision score $\uparrow$ & contact score $\uparrow$ \\
        \midrule
        \model       & 0.894 & 1      \\
        w/o penetration loss  & 0.656 &  1     \\
        w/o contact loss  & 0.995 &    0.194   \\
         \bottomrule
    \end{tabular}
    \label{tab:ablation}
\end{table}

\section{Conclusion}
\looseness=-1

We propose \textbf{S}cene Synthesis from H\textbf{UM}an \textbf{M}oti\textbf{ON}~(\model), a framework that generates multi-object scenes from a sequence of human interaction. \model leverages human contact estimations and scene priors to produce scenes that realistically support the interaction and the semantic context. The flexibility of \model also enables the synthesis of diverse scenes from a single motion sequence. We hope this can also shed light on generating inexpensive diverse human-scene interaction datasets.
In the future, we are interested in exploring the following directions. Since PROXD does not consider soft-body interactions, a potential direction would be considering soft-body deformation of objects such as beds and sofas. Our method considers synthesized scenes to be stationary, hence future works can include movement and rearrangement of furniture during human-scene interaction. As PROXD categorizes all the smaller interaction objects such as books, cups, or TV remotes into a single category, one potential extension to our method would be to include interactions with more specific small objects.


\clearpage
\begin{acks}
This work is in part supported by the Stanford Human-Centered AI Institute (HAI), the Toyota Research Institute (TRI), Innopeak, Meta, Bosch, and Samsung.
\end{acks}

\bibliographystyle{ACM-Reference-Format}
\bibliography{references}
\clearpage

\end{document}